\documentclass[12pt]{article}
\usepackage[pctex32]{graphics}
\begin{document}
\vspace{2cm}
\begin{center}
{\bf  \Large  Quantum Field Effect on Symmetry Breaking and Restoration in Anisotropic Spacetimes }

\vspace{1cm}
                      Wung-Hong Huang\\
                       Department of Physics\\
                       National Cheng Kung University\\
                       Tainan,70101,Taiwan\\
\end{center}
\vspace{2cm}
   The one-loop effective potential for $\phi ^4$ theory on a Bianchi type-I universe is evaluated in the adiabatic approximation. It is used to see the quantum-field effects on symmetry breaking and restoration in anisotropic spacetimes. The results show that the fate of symmetry will not be changed in the cases of conformal coupling or a vanishing scale curvature, and only for some suitable values of scalar-gravitational coupling could the symmetry be radiatively broken or restored. 

\vspace{3cm}
\begin{flushleft}
E-mail:  whhwung@mail.ncku.edu.tw\\
Physical Review D42 (1990) 1287-1288
\end{flushleft}
\newpage
     Quantum fields in curved space have been investigated by many authors [1],  It has been found that gravitational effects play an important role in cosmological phase transition [2-11]. From analyses based on the loop-correction effective potential they then conclude that the scalar-gravitational coupling $\xi$ and the magnitude of the scalar curvature $R$ crucially determine the fate of symmetry.

  The effect of anisotropy in static spacetimes such as mixmaster or Taub universes on the process of symmetry breaking and restoration has been discussed [4,5].   Recently, the Coleman-Weinberg mechanism [12] of symmetry breaking in a Bianchi type-I universe has also been investigated [6].  In Ref.6, the radiative-correction terms arising from the anisotropy are found to be positive and proportional to the magnitude of anisotropy Q. Thus symmetry will be restored for a high-Q value, irrespective of the magnitude of the scalar curvature and value of the scalar-gravitational coupling.  As the effective potential presented in the Ref. 6 is calculated by the perturbation with respect to Q, the conclusion for the case of a large-Q value is untruthful. In this paper we will display the one-loop effective potential for $\phi ^4$ theory in a Bianchi type-I universe in the adiabatic approximation (AA) scheme. Although the AA scheme is valid only in the quasiclassical or WKB region, i.e., a spacetime with slowly varying background fields [13,14],  the value of Q is, however, not limited. (Note: It is not clear that the systems evolving in a spacetime with a large shear Q do not invalidate the slowly varying background field assumption in the treatment of symmetry behavior in dynamical spacetimes [9].) We are therefore able to see the quantum-field effect on the fate of symmetry in a universe with large anisotropy. Our result, however, does not agree with that in Ref. 6. We find that only for some suitable values of scalar-gravitational coupling could the symmetry be radiatively restored, even if the anisotropy is very large.

   The derivative of a one-loop effective potential for massive $\phi^4$ theory in spatially flat Robertson-Walker spacetime has been calculated in Ref.10 [Eq.(39)] by using the AA scheme. It is 

$${\partial V\over \partial \phi_c}= (m_r^2 +\delta m^2 ) \phi_c +(\xi_r +\delta \xi ) R \phi_c + {\lambda_r +\delta \lambda \over 3!}  \phi_c +  {\lambda_r  \over 16\pi^2}  \Lambda^2 \phi_c  ~~~~~~~~~~~~~~~~~~~~~~~~~$$
$${\lambda_r  \phi_c \over 32\pi^2}\left[ m_r^2  + \left(\xi_r -{1\over 6}\right)R + {\lambda_r \over 2}  \phi_c^2  \right] \left[1+ ln\left[{m_r^2 + \left(\xi_r -{1\over 6}\right)R + {\lambda_r \over 2}  \phi_c ^2\over 4 \Lambda^2} \right] \right],  \eqno{(1)} $$
\\
in which the counterterms $\delta m^2$, $\delta\xi$,  and $\delta \lambda$ are needed for renormalization and a momentum cutoff $\Lambda$ is introduced for
regularization. We are interested in the massless $\lambda\phi^4$  theory in Bianchi type-I spacetime. Through the same procedure one can easily show that the derivative of a one-loop effective potential for the theory is just Eq. (1) if the mass $m^2$ is re placed by $Q/a^2$, in which $a^2$ is the average cosmic scale function and $Q$ the anisotropy in the Bianchi type-I spacetime [for the definitions of $a^2$  and $Q$ see Eqs. (5.79) and (5.83) in Reft. 1].

  After introducing the renormalization conditions

      $$0 = {\partial^2 V\over \partial \phi ^2}|_{\phi_c=R=0},~~ \xi_r = {\partial^3 V \over \partial \phi ^2 \partial R}|_{\phi_c=R=0}, ~~\lambda_r = {\partial^4 V\over \partial \phi ^4}|_{\phi_c=R=0}              \eqno{(2)}$$
\\
we finally obtain the derivative of the one-loop renormalized effective potential
                                                                                                                                        $${\partial V\over \partial \phi_c}= \left [\xi_r - {\lambda_r\over 32\pi^2}(\xi_r-{1\over6})\right] R \phi_c + \left[ {\lambda_r \over 3!} -{\lambda_r^2\over 64\pi^2}\right] \phi_c^3 ~~~~~~~~~~~~~~~~~~~~~~~~~~~~~$$
$$ + {\lambda_r  \phi_c \over 32\pi^2}\left[ {\lambda_r\over2} \phi_c^2 + (\xi_r -{1\over 6})R + {Q\over a^2} \right]  ln\left[\lambda \phi_c^2 /2 + (\xi_r-{1\over 6})R + Q/a^2 \over 4 Q/a^2 \right].  \eqno{(3)} $$
\\
Now we are in a position to investigate the gravitational effect on the effective potential in (3). This can be done by considering the case at $\phi \rightarrow 0$; then

$${\partial V\over \partial \phi_c} \rightarrow \xi_r R \phi_c + {\lambda_r  \phi_c \over 32\pi^2} \left[ - (\xi_r -{1\over 6})R  + \left[(\xi_r -{1\over 6})R + {Q\over a^2}\right]  ln\left[(\xi_r -{1\over 6})R + Q/a^2 \over  Q/a^2 \right]\right].  \eqno{(4)} $$
\\
The second term, which is the radiative correction term, is zero if  $\xi_r = {1\over6}$ and/or $R =0$. Thus, in the cases of conformal coupling or vanishing scalar curvature the one-loop quantum correction does not change the fate of symmetry of the tree level.  For the other cases, we can from the above equation find that only for some suitable values of scalar-gravitational coupling could the symmetry be radiatively broken or restored.

\newpage
\begin{enumerate}
\item  N. D. Birrell and P. C. W. Davies, Quantum Fields in Curved   Space (Cambridge University Press, Cambridge, England, 1982).
\item  B. L. Hu, in Proceedings of the Tenth International Conference  on General Relativity and Gravitation, Padua, Italy, 1983,  edited by B. Bertotti, F. deFelice, and A. Pascolini (Consiglio Nazionale Delle Ricerche, Roma, 1983), p. 1086.
\item  L. F. Abbott, Nucl. Phys. B185, 233 (1981); L. H. Ford and D.  J. Toms, Phys. Rev. D 25, 1510 (1982); B. Allen, Nucl. Phys. B226, 228 (1983); Ann. Phys. (N.Y.) 161, 152 (1983); A. Vilenkin, Nucl. Phys. B226, 504 (1983).
\item  R. Critchley and J. S. Dowker, J. Phys. A 15, 157 (1982).
\item T. C. Shen, B. L. Hu, and D. J. O'Connor, Phys. Rev. D 31,
  2401 (1985).
\item  T. Futamase, Phys. Rev. D 29, 2789 (1984).
\item  B. L. Hu and D. J. O'Connor, Phys. Rev. D 30, 74 (1984); 34, 2535 (1987); 36, 1701 (1987); B. L. Hu and Y. Zhang, ibid. 37, 2125 (1988).
\item  B. L. Hu, Phys. Lett. 123B, 189 (1983); L. F. Chen and B. L. Hu, ibid. 160B, 36 (1985).
\item  S. Sinha and B. L. Hu, Phys. Rev. D 38, 2423 (1988).
\item  A. Ringwald, Phys. Rev. D 36, 2598 (1987).
\item A. Ringwald, Ann. Phys. (N.Y.) 177, 129 (1987).
\item S. Coleman and E. Weinberg, Phys. Rev. D 7, 1888 (1973).
\item  Ya. B. Zel'dovich and A. A. Starobinsky, Zh. Eksp. Teor. Fiz. 61, 2162 (1972) [Sov. Phys. JETP 34, 1159 (1972)].
\item  L. Parker and S. A. Fulling, Phys. Rev. D 9, 341 (1974); S. A. Fulling, L. Parker, and B. L. Hu, ibid. 10, 3905 (1974).
\end{enumerate}
\end{document}